\documentclass[12pt]{article}
\usepackage{epsfig}
\usepackage{citesort}
\usepackage{amssymb,amsfonts}
\newcommand{\beq}{\begin{equation}}
\newcommand{\eeq}{\end{equation}}
\newcommand{\bea}{\begin{eqnarray}}
\newcommand{\eea}{\end{eqnarray}}

\begin{document}
\title{\bf Phase diagram and thermodynamics\\ of the PNJL model\footnote{Work supported in part by BMBF and GSI}}
\author{Claudia Ratti$^{a,b}$, Michael A. Thaler$^a$ and Wolfram Weise$^a$ 
\\\small{$^a$Physik-Department, Technische Universit\"at M\"unchen, D-85747 
Garching, Germany}\\ 
\small{$^b$ECT$^*$, I-38050 Villazzano (Trento), Italy}\\ 
\small{and INFN,
Gruppo Collegato di Trento, Via Sommarive, 38050 POVO
}\\}
\date{\today}
\maketitle
\begin{abstract}
QCD-based thermodynamics at zero and finite quark chemical
potential is studied using an extended Nambu and Jona-Lasinio approach
in which quarks couple simultaneously to the chiral condensate and to a background temporal 
gauge field representing Polyakov loop dynamics. This so-called PNJL model thus
includes features of both deconfinement and chiral symmetry restoration. We discuss the
phase diagram as it emerges from this approach in close comparison with results from lattice QCD thermodynamics. The critical point, separating
crossover from first order phase transition, is investigated with special focus on its quark mass dependence, starting from the relatively large masses presently accessible
by lattice simulations, down to the chiral limit.
\end{abstract}

\section{Introduction and Basics}

Models of the Nambu and Jona-Lasinio (NJL) type \cite{NJL61} have a long history and have been
used extensively to describe the dynamics and thermodynamics of the lightest hadrons \cite{VW91,Kl92,HK94,Ri97}, including investigations of phase diagrams \cite{Bu05,Barducci:2005ut}. Such schematic models offer a simple and practical  
illustration of the basic mechanisms that drive spontaneous chiral symmetry breaking, a key feature
of QCD in its low-temperature, low-density phase. 

The NJL model is based on an effective Lagrangian of relativistic fermions (quarks) which interact
 through local current-current couplings, assuming that gluonic degrees of freedom can be frozen into pointlike effective interactions between quarks. Lattice QCD results for the gluonic field strength correlation function \cite{DiG} demonstrate that the colour correlation length, i.e. the distance over which colour fields propagate in the QCD vacuum, is small, of order $0.2$ fm corresponding to a characteristic momentum scale $\Lambda$ of order 1 GeV.  Consider now the basic non-local interaction between two quark colour currents, $J_i^\mu = \bar{\psi}\gamma^\mu t_i\psi$, where $t_i$ are the generators of the $SU(N_c)$ colour gauge group. The contribution of this current-current coupling to 
the action is:
\begin{equation}
S_{int} = -{1\over 2} \int d^4x \,d^4y\,J_i^\mu(x) \,g^2D_{\mu\nu}^{ij}(x,y)\,J_j^\mu(y)~~,
\label{action}
\end{equation}
where $D_{\mu\nu}^{ij}$ is the full gluon propagator and $g$ is the QCD coupling. In perturbative QCD this $S_{int}$ generates the familiar one-gluon exchange interaction between quarks and maintains its non-local structure. That is the situation realised in the quark-gluon phase at extremely high temperatures.  As one approaches the hadronic phase around a critical temperature of about 0.2 GeV, the propagating gluons experience strong screening effects which cannot be handled perturbatively.  If the range over which colour can be transported is now restricted to the short distance scale $\Lambda^{-1}$, while typical momentum scales (Fermi momenta) of the quarks are small compared to $\Lambda$, then the quarks experience an interaction which can be approximated by a local coupling between their colour currents:
\begin{equation}
{\cal L}_{int} = -G_c\,J_\mu^i(x)\,J^\mu_i(x)\,\, ,
\label{Lint}
\end{equation}
where $G_c \sim \bar{g}^2\, \Lambda^{-2}$ is an effective coupling strength of dimension $length^2$ which encodes the QCD coupling, averaged over the relevant distance scales, in combination with the squared correlation length, $\Lambda^{-2}$. 
In essence, by ``integrating out" gluon degrees of freedom and absorbing them in the four-fermion interaction ${\cal L}_{int}$, the local $SU(N_c)$ gauge symmetry of QCD is now replaced by a global $SU(N_c)$ symmetry of the NJL model.  Apart from this step, the interaction Lagrangian (\ref{Lint}) evidently preserves the chiral $SU(N_f) \times SU(N_f)$ symmetry that it shares with the original QCD Lagrangian for $N_f$ massless quark flavours.

A Fierz transform of the colour current-current interaction (\ref{Lint}) produces a set of exchange terms
acting in quark-antiquark channels. For the $N_f = 2$ case: 
\begin{equation}
{\cal L}_{int} \rightarrow {G\over 2}\left[(\bar{\psi}\psi)^2 + (\bar{\psi}i\gamma_5\vec{\tau}\psi)^2\right] + ... \,\, , 
\label{njl}
\end{equation}
where $\vec{\tau} = (\tau_1,\tau_2, \tau_3)$ are the isospin $SU(2)$ Pauli matrices. Not shown for brevity is a series of terms with combinations of vector and axial vector currents, both in colour singlet and colour octet channels. The constant $G$ is proportional to the colour coupling strength $G_c$. Their ratio is uniquely determined by $N_c$ and $N_f$.

Eq.(\ref{njl}) is the starting point of the standard NJL model. In the mean field (Hartree) approximation, the NJL equation of motion leads to the gap equation
\begin{equation}
m = m_0 - G\langle\bar{\psi}\psi\rangle\,\, .
\end{equation}
With a small bare (current) quark mass $m_0$ as input,
this equation links the dynamical generation of a large constituent quark mass $m$ to spontaneous chiral symmetry breaking and the appearance of the quark condensate
\begin{equation}
\langle\bar{\psi}\psi\rangle= -Tr\lim_{\,x\rightarrow \,0^+}\langle {\cal T}\psi(0)\bar{\psi}(x)\rangle = -2iN_fN_c\int {d^4p\over (2\pi)^4}{m\,\,\theta(\Lambda^2 -\vec{p}\,^2)\over p^2 - m^2 + i\varepsilon}\,\, .
\end{equation}
For $m_0 = 0$ a non-zero quasiparticle mass develops dynamically, together with a non-vanishing chiral condensate, once $G$ exceeds a critical value. The procedure requires a momentum cutoff $\Lambda \simeq 2m$ beyond which the interaction is ``turned off". Note that the strong interaction, by polarizing the vacuum and turning it into a condensate of quark-antiquark pairs, transforms an initially pointlike quark with its small bare mass $m_0$ into a massive quasiparticle with a finite size. (Such an NJL-type mechanism is commonly thought to be at the origin of the phenomenological constituent quark masses $m \sim$ 0.3-0.4 GeV). 

While the NJL model illustrates the transmutation of  originally light (or even massless) quarks and antiquarks into massive quasiparticles, it generates at the same time pions as Goldstone bosons of spontaneously broken chiral symmetry. NJL type approaches have also been used extensively to explore colour superconducting phases at high densities through the formation of various sorts of diquark condensates \cite{ARW98,Bu05}.

Despite their widespread use, NJL models have a principal deficiency. The
reduction to global (rather than local) colour symmetry has the consequence
that quark confinement is missing. Confinement is the second key feature of
low-energy QCD besides spontaneous chiral symmetry breaking. While confinement
is a less significant aspect for $N_c=2$ thermodynamics which can be described
quite successfully using the simplest NJL approach~\cite{Ratti}, it figures
prominently for $N_c=3$ QCD. There have been ongoing discussions whether 
deconfinement and the restoration of chiral symmetry are directly connected in 
the sense that they appear at the same transition temperature $T_c$, as 
suggested by 
earlier lattice computations. In any case, as one approaches $T_c$ from above, 
all versions of the ``classic" NJL model encounter the problem that they 
operate with the ``wrong" degrees of freedom. Quarks as coloured 
quasiparticles are incorrectly permitted to propagate over large distances 
even in the hadronic sector of the phase diagram. In contrast, confinement and 
spontaneous chiral symmetry breaking imply that QCD below $T_c$ turns into a 
low-energy effective theory of weakly interacting Goldstone bosons (pions) 
with derivative couplings to colour-singlet hadrons (rather than quarks). 

In the limit of infinitely heavy quarks, the deconfinement phase transition is characterized by spontaneous breaking of the $Z(3)$ center symmetry of QCD. The corresponding order parameter is the thermal Wilson line, or Polyakov loop, winding around the imaginary time direction with periodic boundary conditions:
\beq
\phi\left(\vec{x}\right)=N_c^{-1}\,Tr \,\mathcal{P}\exp\left[i\int_{0}^{\beta}d\tau\,A_4\left(\vec{x},\tau\right)\right],
\label{polyakov}
\eeq
with $\beta = 1/T$ the inverse temperature. Here $A_4 = i A^0$ is the temporal component of the Euclidean gauge field $(\vec{A}, A_4)$ and  $\mathcal{P}$ denotes path ordering. In the presence of dynamical quarks the $Z(3)$ symmetry is explicitly broken. The Polyakov loop ceases to be a rigourous order parameter but still serves as an indicator of a rapid crossover towards deconfinement.

Recent developments have aimed at a synthesis of the NJL model with Polyakov loop dynamics. The principal idea is to introduce both the chiral condensate $\langle\bar{\psi}\psi\rangle$ and the Polaykov loop $\phi$ as classical, homogeneous fields which couple to the quarks according to rules dictated by the symmetries and symmetry breaking patterns of QCD, thus unifying aspects of confinement and chiral symmetry breaking. We refer to this combined scheme as the PNJL (Polyakov-loop-extended NJL) model. The present writeup is largely based on our previous ref.\cite{noi}. It reviews results of calculations as well as presenting an outlook for further steps to be pursued in close comparison with results from lattice QCD thermodynamics. 

\section{Introducing the PNJL model}
Throughout this presentation we work with two flavours $(N_f = 2)$ and specify the PNJL Lagrangian \cite{noi} as follows. Its basic ingredients are the Nambu and Jona-Lasinio type four-fermion contact term and the
coupling to a (spatially constant) temporal background gauge field representing Polyakov loop dynamics:
\bea
\mathcal{L}_{PNJL}=\bar{\psi}\left(i\gamma_{\mu}D^{\mu}-\hat{m}_0
\right)\psi+\frac
{G}{2}\left[\left(\bar{\psi}\psi\right)^2+\left(\bar{\psi}i\gamma_5
\vec{\tau}\psi
\right)^2\right]
-\mathcal{U}\left(\phi[A],\phi^*[A];T\right),
\label{lagr}
\eea
where $\psi=\left(\psi_u,\psi_d\right)^T$ is the quark field,
\bea
D^{\mu}=\partial^\mu-i A^\mu~~~~~~\mathrm{and}~~~~~~A^\mu=\delta^{\mu}_{0}A^0~~,
\eea
with $A^0 = -iA_4$. The gauge coupling $g$ is conveniently absorbed in the definition of $A^\mu(x) = g {\cal A}^\mu_a(x){\lambda_a\over 2}$ where ${\cal A}^\mu_a$ is the SU(3) gauge field and $\lambda_a$ are the Gell-Mann matrices. The two-flavour current quark mass matrix is $\hat{m}_0 = diag(m_u, m_d)$ and we shall work in the isospin symmetric limit with $m_u = m_d \equiv m_0$. As previously mentioned, $G$ is the coupling strength of the chirally symmetric four-fermion interaction. 

The effective potential $\mathcal{U}(\phi,\phi^*;T)$ is  expressed in 
terms of the traced Polyakov loop (\ref{polyakov}), reduced to our case of a constant Euclidean field $A_4$:
\beq
\phi={1\over 3}\mathrm{Tr}_c\,\exp\left[{iA_4\over T}\right]~~~.
\eeq
In a convenient gauge (the so-called Polyakov gauge), the Polyakov loop matrix can be given a  
diagonal representation~\cite{Fukushima:2003fw}.

The effective potential $\mathcal{U}$ has the following general features.
It must satisfy the $Z(3)$ center symmetry just like the pure gauge QCD Lagrangian.
Furthermore, in accordance with lattice results for the behaviour of the
Polyakov loop as a function of temperature $T$, 
the potential ${\cal U}$ must have a single minimum at $\phi$ = 0 at small $T$, while at high $T$ it develops a second minimum which becomes the absolute minimum
above a critical temperature $T_0$. In the limit $T\rightarrow\infty$ we have
$\phi\rightarrow~1$. The following general form is chosen for  ${\cal U}$, including a $\phi^3$ term which reflects the underlying $Z(3)$ symmetry:
\bea
{\mathcal{U}\left(\phi,\phi^*;T\right)\over T^4} =-{b_2\left(T\right)\over 2}\phi^* \phi-
{b_3\over 6}\left(\phi^3+
{\phi^*}^3\right)+ {b_4\over 4}\left(\phi^* \phi\right)^2
\label{u1}
\eea
with
\beq
b_2\left(T\right)=a_0+a_1\left(\frac{T_0}{T}\right)+a_2\left(\frac{T_0}{T}
\right)^2+a_3\left(\frac{T_0}{T}\right)^3~~~.
\label{u2}
\eeq
A precision fit of the coefficients $a_i,~b_i$ is performed to reproduce the pure-gauge lattice data.

There is a subtlety about the Polyakov loop field, $\phi$, and its conjugate, $\phi^*$, in the presence of quarks. At  zero chemical potential we have $\phi=\phi^*$, i.e. the field $\phi$ is real, it serves as an order parameter for deconfinement and a mean-field calculation is straightforward. At non-zero quark chemical potential, $Z(3)$ symmetry is explicitly broken and $\phi$ differs from $\phi^*$ while their thermal expectation values $\langle\phi\rangle$ and $\langle\phi^*\rangle$ remain real \cite{DPZ05}. A detailed analysis of the stationary points of the action under these conditions requires calculations beyond mean field which will be reported elsewhere \cite{RRW}. We proceed here, as in \cite{noi},  by introducing $\Phi \equiv  \langle\phi\rangle$ and $\bar{\Phi} \equiv \langle\phi^*\rangle$ as new independent field variables which replace $\phi$ and $\phi^*$ in Eq.(\ref{u1}). This approximate prescription  corresponds to a modified mean-field scheme which can account for the difference between $\Phi$ and $\bar{\Phi}$ in the presence of quarks. The more accurate treatment is under way.  

Using standard bosonization techniques, we introduce 
the auxiliary bosonic fields $\sigma$ and $\vec{\pi}$ for the scalar-isoscalar and pseudoscalar-isovector
quark bilinears in Eq.(\ref{lagr}).
The expectation value of the $\sigma$ field is directly related to the
chiral condensate by  $\langle\sigma\rangle=G\langle\bar{\psi}\psi\rangle$ and the gap equation becomes
\beq
m=m_0-\langle\sigma\rangle~~.
\label{mass}
\eeq
Note that $\langle\sigma\rangle$ is negative in
our representation, and the chiral (quark) condensate is $\langle\bar{\psi}\psi\rangle=\langle\bar{\psi}_u
\psi_u+\bar{\psi}_d\psi_d\rangle$.

Before passing to the actual calculations, we summarize basic 
assumptions behind Eq.(\ref{lagr}) and comment on limitations to be kept in 
mind. The PNJL model reduces 
gluon dynamics to a) chiral point couplings between quarks, and b) a simple
static background field representing the Polyakov loop. This
picture can be expected to work only within a limited range of temperatures.
At large $T$, transverse gluons are known to be thermodynamically active 
degrees of freedom, but they are ignored in the PNJL model. To what extent this
model can reproduce lattice QCD thermodynamics is nonetheless a relevant 
question. We can assume that its range of applicability is, roughly, $T\leq (2-3)T_c$, based
on the conclusion drawn in ref.~\cite{Meisinger:2003id} that transverse gluons
start to contribute significantly for $T>2.5\,T_c$. 

\section{Parameter fixing\label{sec3}}
\begin{figure}
\begin{minipage}[t]{.48\textwidth}
\parbox{5cm}{
\scalebox{.78}{
\includegraphics*[59,522][381,730]{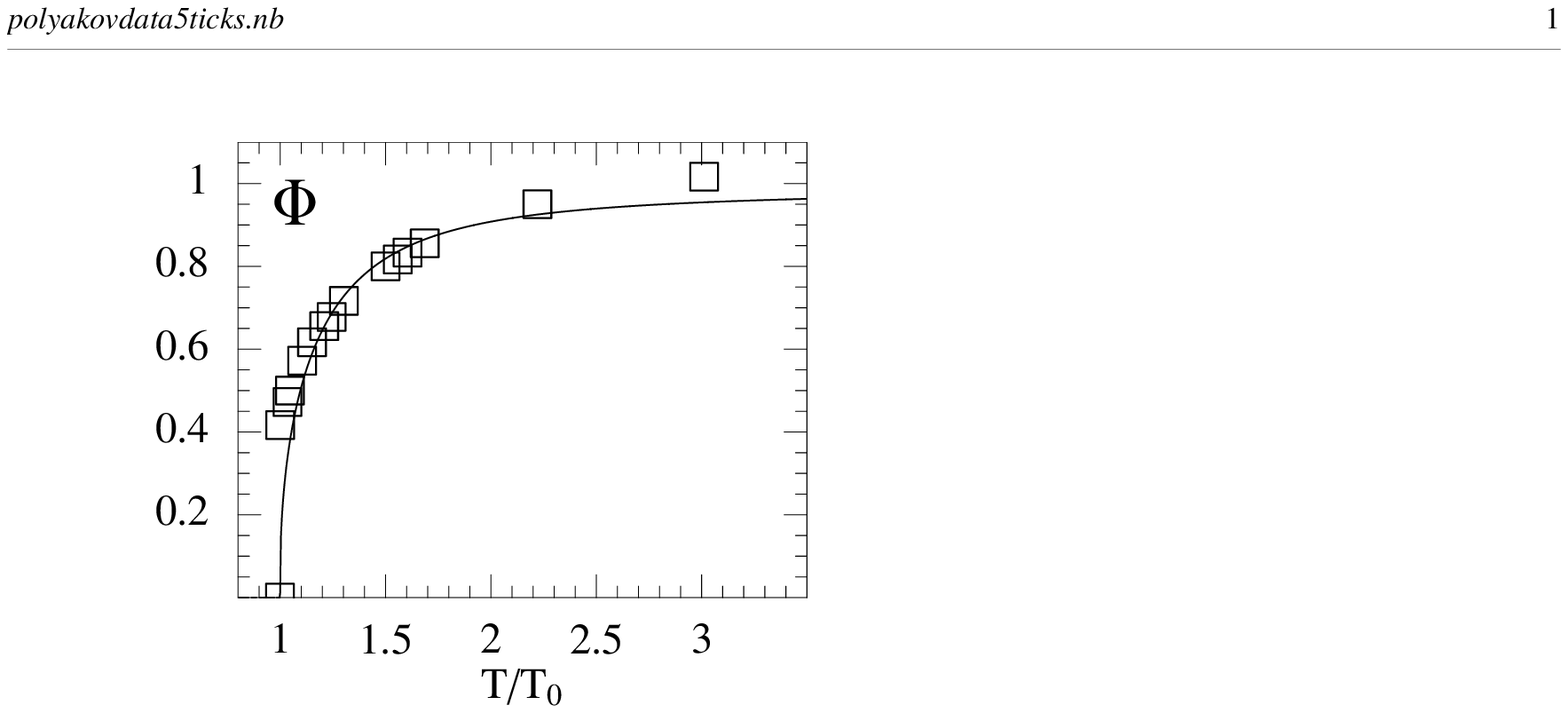}}
}\\
\centerline{(a)}
\end{minipage}
\begin{minipage}[t]{.48\textwidth}
\parbox{5cm}{
\scalebox{.75}{
\includegraphics*[59,502][381,730]{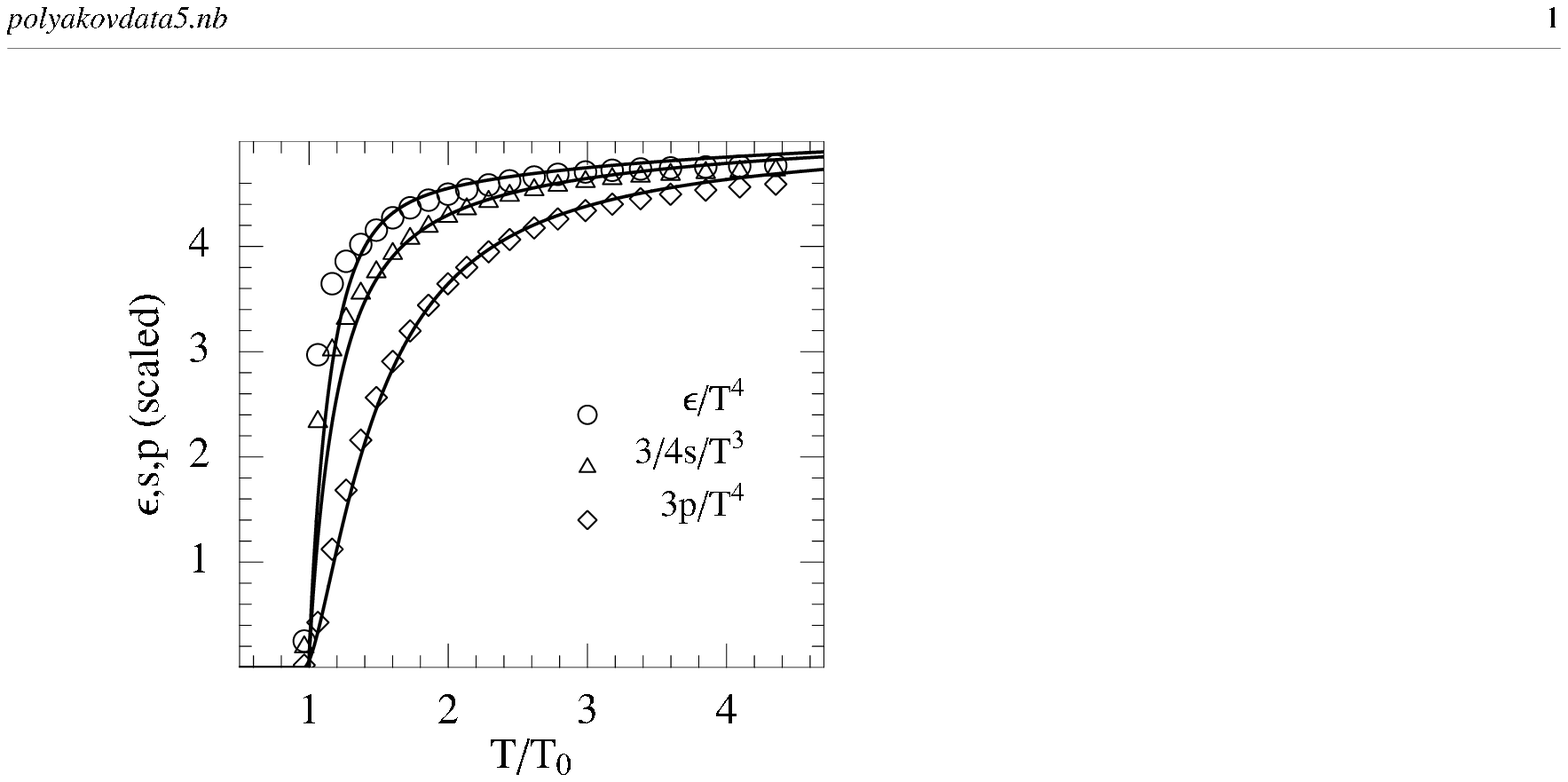}}}\\
\centerline{(b)}
\end{minipage}
\parbox{15cm}{
\caption{
\footnotesize (a): Polyakov loop as a function of  
temperature in the pure gauge sector, compared to corresponding lattice results
taken from Ref.~\cite{Kaczmarek:2002mc}. (b): Scaled pressure, entropy density and energy density as 
functions
of the temperature in the pure gauge sector, compared to the corresponding 
lattice data taken from Ref.~\cite{Boyd}.
\label{fig1}}}
\end{figure}

The parameters of the Polyakov loop potential $\mathcal{U}$ are fitted to 
reproduce the lattice data~\cite{Boyd} for QCD thermodynamics in the pure gauge sector.
Minimizing $\mathcal{U}(\Phi,\bar{\Phi},T)$ one has $\Phi=\bar{\Phi}$ and the 
pressure of the pure-gauge system is evaluated as $p(T)=-\mathcal{U}(T)$ with $\Phi(T)$ 
determined at the minimum.
The entropy and energy density are then obtained by means of the standard
thermodynamic relations.
Fig.\ref{fig1}(a) shows the behaviour of the Polyakov loop as a function of temperature,
while Fig.\ref{fig1}(b) displays the corresponding
(scaled) pressure, energy density and entropy density. 
The lattice data are
reproduced extremely well using the ansatz~(\ref{u1},\ref{u2}), with 
parameters summarized in Table~\ref{table1}.
The critical temperature $T_0$ for deconfinement appearing in Eq.~(\ref{u2}) is
fixed at $T_0=270$ MeV in the pure gauge sector.
\begin{table}
\begin{center}
\begin{tabular}{|c|c|c|c|c|c|}
\hline
\hline
&&&&&\vspace{-.3 cm}\\
$a_0$&$a_1$&$a_2$&$a_3$&$b_3$&$b_4$\\
&&&&&
\vspace{-.3cm}\\
\hline
\vspace{-.3cm}\\
6.75&-1.95&2.625&-7.44&0.75&7.5\\
\hline
\hline
\end{tabular}
\caption{Parameter set used in \cite{noi} for the Polyakov loop 
potential~(\ref{u1},~\ref{u2}).}
\label{table1}
\end{center}
\end{table}

The pure NJL model part of the Lagrangian~(\ref{lagr}) has the following 
parameters: the ``bare'' quark mass $m_0$, 
a three-momentum cutoff $\Lambda$ and the coupling strength $G$.
We fix them by reproducing the known chiral physics in the hadronic sector at 
$T=0$: the pion decay constant $f_\pi$, the chiral condensate 
$|\langle{\bar \psi}_u\psi_u\rangle|$$^{1/3}$ and the pion mass $m_\pi$ are
evaluated in the model and adjusted at their empirical values.
The results are shown in Table~\ref{table2}.
\begin{table}
\begin{center}
\begin{tabular}{|c|c|c|}
\hline
\hline
&&\vspace{-.3 cm}\\
$\Lambda$ [GeV]&$G$[GeV$^{-2}$]&$m_0$[MeV]\\
&&\vspace{-.3cm}\\
\hline
&&\vspace{-.3cm}\\
0.651&10.08&5.5\\
&&\vspace{-.3cm}\\
\hline
\hline
&&\vspace{-.3cm}\\
$|\langle{\bar \psi}_u\psi_u\rangle|$$^{1/3}$[MeV]&$f_{\pi}$[MeV]&$m_{\pi}$[MeV]\\
&&\vspace{-.3cm}\\
\hline
&&\vspace{-.3cm}\\
251&92.3&139.3\\
\hline
\hline
\end{tabular}
\caption{Parameter set used for the NJL model part of the effective Lagrangian~(\ref{lagr}),
and the resulting physical quantities. These values of the parameters yield a constituent quark mass $m=$ 325 MeV.}
\label{table2}
\end{center}
\end{table}
\section{Finite $\mu$ thermodynamics}
\subsection{General features}
We now extend the model to finite temperature and
chemical potentials using the Matsubara formalism. We
consider the isospin symmetric case, with an equal number of $u$ and $d$ quarks (and therefore a 
single quark chemical potential $\mu$).
The quantity to be minimized at finite temperature is the thermodynamic
potential per unit volume:
\bea
\nonumber
\Omega&=&{\cal U}\left(\Phi,\bar{\Phi},T\right)+\frac{\sigma^2}{2G}\\
&-&2N_f\,T\int\frac{\mathrm{d}^3p}{\left(2\pi\right)^3}
\left\{\mathrm{Tr}_c\ln\left[1+\mathrm{e}^{-\left(E_p-\tilde{\mu}
\right)/T}\right]
+\mathrm{Tr}_c\ln\left[1+\mathrm{e}^{-\left(E_p+
\tilde{\mu}\right)/T}\right]\right\}
\nonumber\\ 
&-& 6N_f\int\frac{\mathrm{d}^3p}{\left(2\pi\right)^3}\,{E_p}\,\,\theta(\Lambda^2-\vec{p}\,^2)~~~,
\label{omega2}
\eea
where $\tilde{\mu} = \mu + iA_4$ and $E_p=\sqrt{\vec{p}\,^2+m^2}$ is the quark quasiparticle energy. The last term involves the NJL three-momentum 
cutoff $\Lambda$. The second (finite) term does not require any cutoff.

Notice that the coupling of the Polyakov loop to quarks effectively reduces the residues at the quark quasiparticle poles as the critical temperature is approached from $T > T_c$: expanding the logarithms in the second line of (\ref{omega2}) one finds $\mathrm{Tr}_c\ln\left(1+\exp[-(E_p-\mu-iA_4)/T]\right) = 3\,\phi\,\exp[-(E_p-\mu)/T] + ...$ , with $\phi$ then to be replaced by $\langle\phi\rangle \equiv \Phi$ which tends to zero as $T\rightarrow T_c$.

From the thermodynamic potential (\ref{omega2}) the equations of motion for the mean fields $\sigma, \Phi$ and $\bar{\Phi}$ are determined through
\beq
{\partial\Omega\over\partial\sigma} = 0 ~,~~~~~{\partial\Omega\over\partial\Phi} = 0 ~,~~~~~{\partial\Omega\over\partial \bar{\Phi}} = 0 ~.
\eeq
This set of coupled equations is then solved for the fields as functions of temperature $T$ and quark chemical potential $\mu$.
\begin{figure}
\hspace{-.05\textwidth}
\begin{minipage}[t]{.48\textwidth}
\includegraphics*[width=\textwidth]{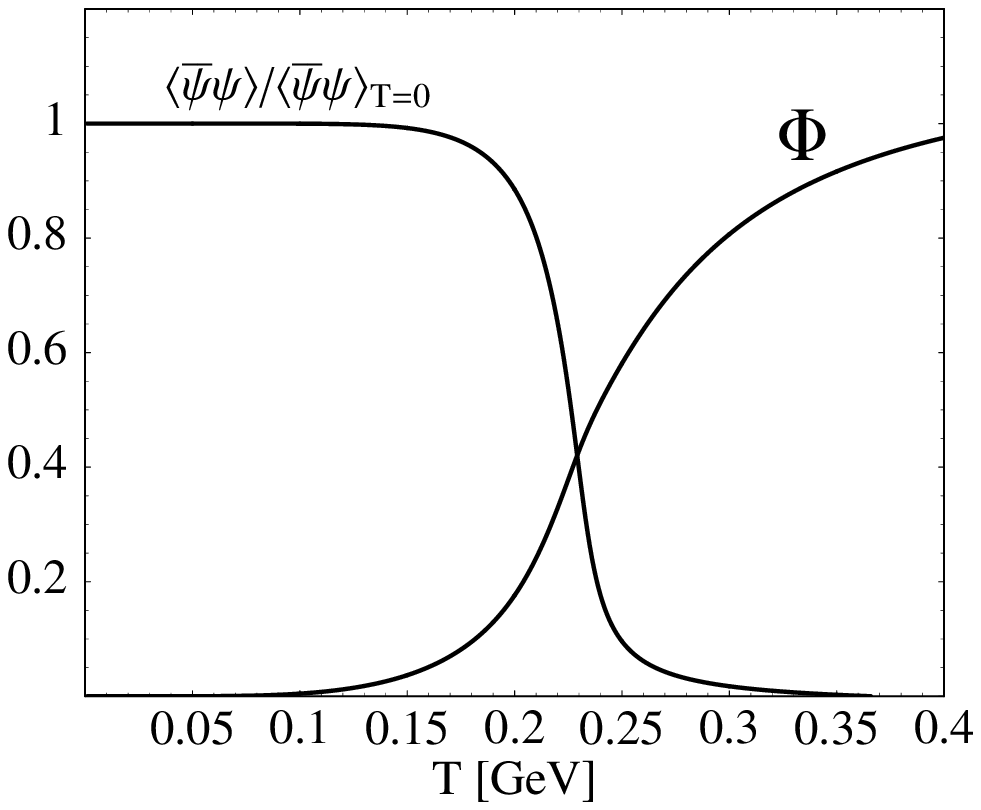}\\
\centerline{(a)}
\end{minipage}
\hspace{-.05\textwidth}
%
\begin{minipage}[t]{.48\textwidth}
\hspace{-.15\textwidth}
\scalebox{.85}{
\includegraphics*[width=1.17\textwidth]{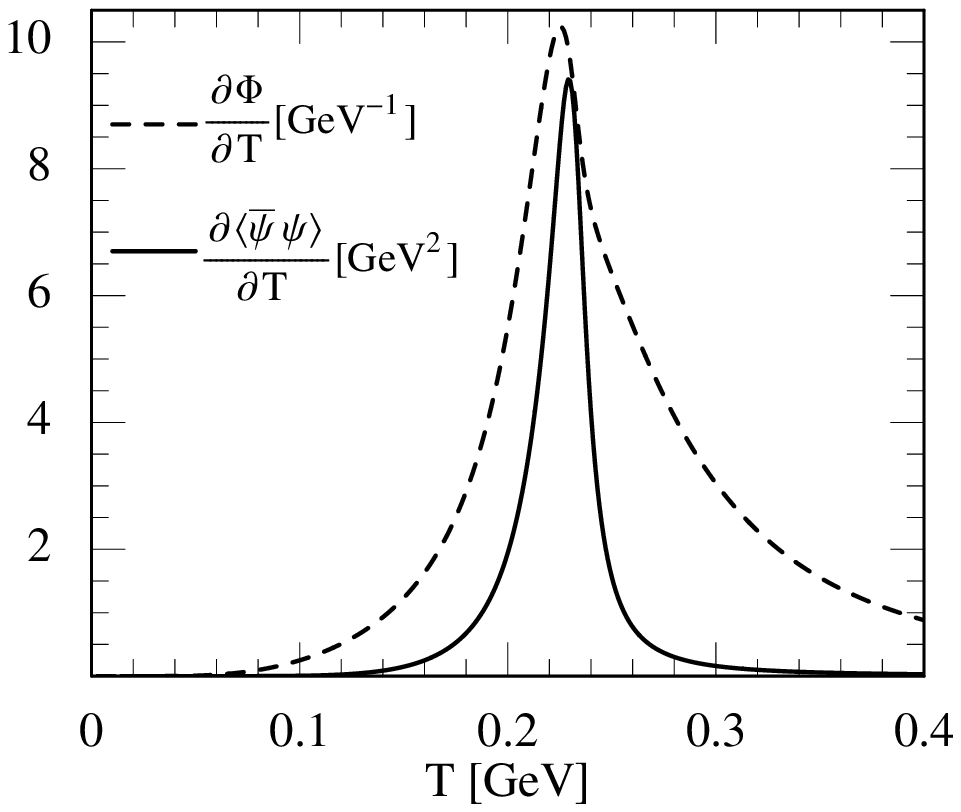}}\\
\centerline{(b)}
\end{minipage}
\parbox{15cm}{
\caption{Left: scaled chiral condensate and Polyakov loop 
$\Phi(T)$ as
functions of temperature at zero chemical potential. Right: plots of
$\partial\langle\bar{\psi}\psi\rangle/\partial T$ and $\partial\Phi/
\partial T$.
}
\label{fig2}}
\end{figure}
Fig.~\ref{fig2}(a) shows the chiral condensate together with the Polyakov loop 
$\Phi$ as functions of temperature at $\mu = 0$ where we find 
$\Phi=\bar{\Phi}$.
One observes that the introduction of quarks coupled to the $\sigma$ and
$\Phi$ fields turns the first-order transition seen in pure-gauge lattice QCD
into a continuous crossover. The 
crossover transitions for the chiral condensate 
$\langle\bar{\psi}\psi\rangle$ and for the Polyakov loop almost coincide 
at a critical temperature $T_c \simeq 220$ MeV (see Fig.~\ref{fig2}(b)). 
We point out that this feature is obtained without changing a single
parameter with respect to the pure gauge case.
The value of the critical temperature found here is a little high if 
compared to the available data for two-flavour Lattice QCD 
\cite{Karsch} which give $T_c = (173 \pm 8)$ MeV.
For quantitative comparison with existing lattice results we choose to reduce 
$T_c$ by rescaling the parameter $T_0$ from 270 to 190 MeV. In this case we 
loose the perfect
coincidence of the chiral and deconfinement transitions, but they are shifted
relative to each other by less than 20 MeV. When defining $T_c$ in this case
as the average of the two transition temperatures we find $T_c=180$ MeV.

\subsection{Detailed comparison with lattice data}
\begin{figure}
\hspace{.2\textwidth}
\parbox{5cm}{
\scalebox{.95}{
\includegraphics*[width=.55\textwidth]{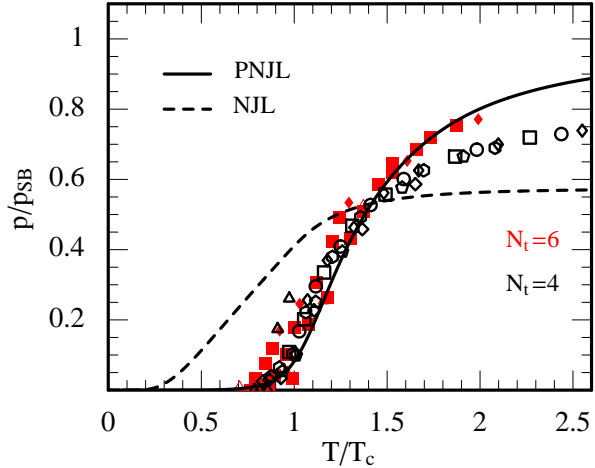}\\}}

%
%
\parbox{15cm}{
\caption{
\footnotesize Scaled pressure divided by the Stefan-Boltzmann (ideal gas) 
limit as a function of temperature at zero chemical
potential: comparison between our PNJL model prediction \cite{noi} (full line), the 
standard NJL model result (dashed) and 
lattice results corresponding to $N_t=4$ and $N_t=6$. Lattice data are taken 
from Ref.~\cite{AliKhan:2001ek}}
\label{fig4}}
\end{figure}

The primary aim is now to compare predictions of our PNJL model with 
the lattice data available for full QCD thermodynamics (with quarks included) at zero and finite 
chemical potential $\mu$. Consider first the pressure $p\left(T,\mu=0\right)=-\Omega\left(T,\mu=0\right)$ of the quark-gluon system at zero chemical potential.
Our results are presented in Fig.\,\ref{fig4} in comparison with 
corresponding lattice data. We point out that the input parameters
of the PNJL model have been fixed independently in the pure gauge and hadronic 
sectors, so that the calculated pressure is a prediction of the model, without
any further tuning of parameters. 
With this in mind, the agreement with lattice results is quite satisfactory.
Also shown in Fig.\,\ref{fig4} is the result obtained in the standard NJL model.
Its deficiencies are evident. At low temperatures the pressure comes out incorrect. The missing
confinement permits quarks to be active degrees of freedom even in the forbidden region $T<T_c$. 
At high temperatures, the standard NJL result for the pressure is significantly lower than the one
 seen in the lattice data. The gluonic thermodynamics is missing altogether in the NJL model, 
whereas in the PNJL model it is partially taken into account by means of the 
Polyakov loop effective potential $\mathcal{U}(\Phi,\bar{\Phi},T)$. As stated previously,
the range of validity of this approach is limited, however,  to temperatures smaller 
than $2.5~T_c$, beyond which transverse gluon degrees of freedom become important.

The introduction of the Polyakov loop within the PNJL quasiparticle model leads to a 
remarkable improvement in basically all thermodynamic quantities.
The coupling of the quark quasiparticles 
to the field $\Phi$ reduces their weight as thermodynamically active degrees of
freedom when the critical temperature $T_c$ is approached from above. The quasiparticle exponentials
exp$[-(E_p \pm \mu)/T]$ are progressively suppressed in the thermodynamic 
potential as $T \rightarrow T_c$. This is what can be interpreted as the effect
of confinement in the context of the PNJL model.

One must note that the lattice data are grouped in different sets obtained on 
lattices with temporal extent $N_t = 4$ and $N_t = 6$, both of which are not 
continuum extrapolated. In contrast, our calculation should, strictly speaking,
be compared to the continuum limit. In order to perform meaningful comparisons,
the pressure is divided by its asymptotic high-temperature (Stefan-Boltzmann) 
limit for each given case. At high temperatures our predicted curve should be 
located closer to the $N_t = 6$ set than to the one with $N_t = 4$. This is 
indeed the case.

\begin{figure}
\hspace{-.05\textwidth}
\begin{minipage}[t]{.48\textwidth}
\includegraphics*[width=\textwidth]{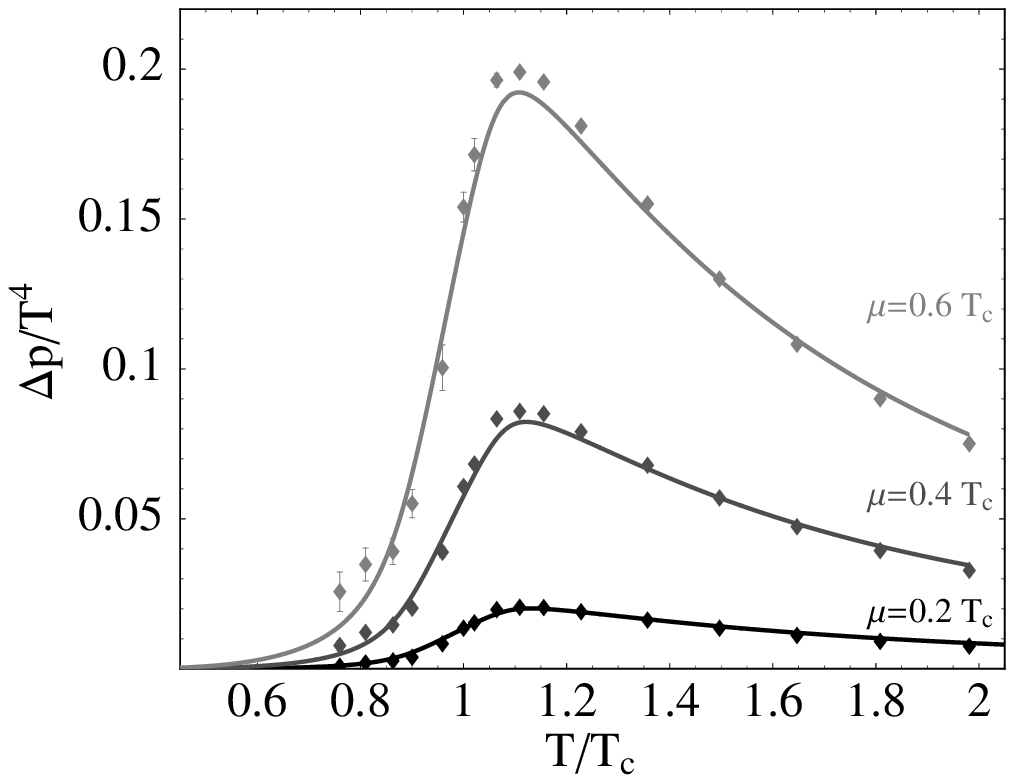}\\
\centerline{(a)}
\end{minipage}
\hspace{.02\textwidth}
\begin{minipage}[t]{.48\textwidth}
\includegraphics*[width=\textwidth]{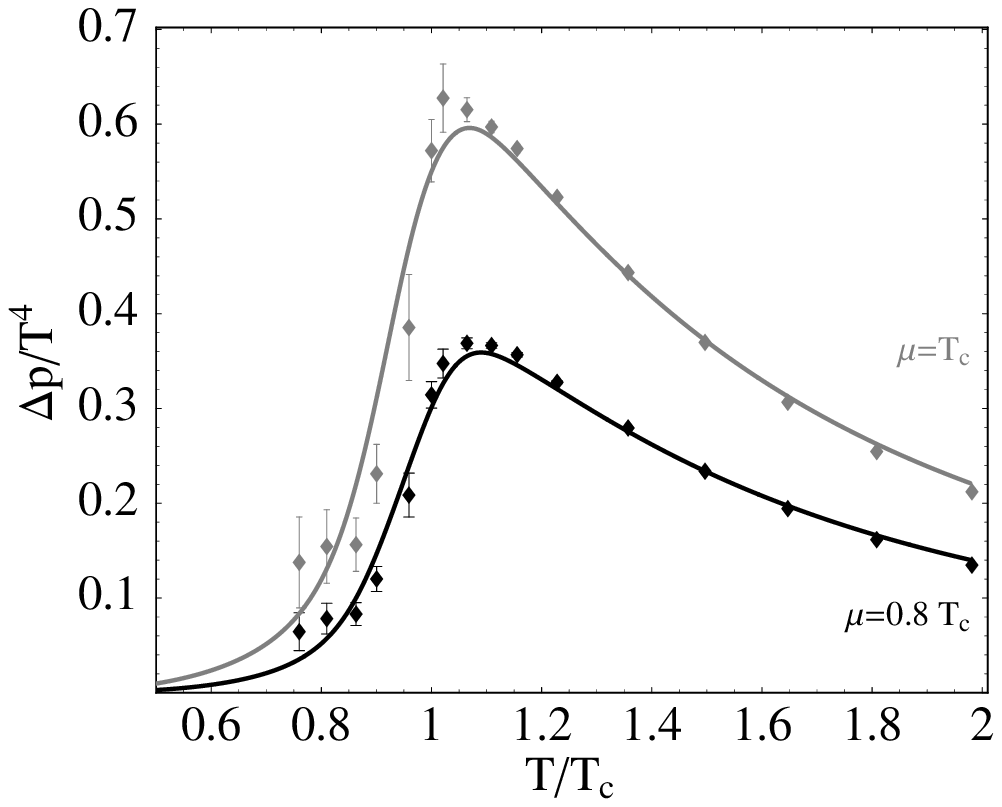}\\
\centerline{(b)}
\end{minipage}
\parbox{15cm}{
\caption{
\footnotesize Scaled pressure difference as a function of temperature at
different values of the quark chemical potential (results from Ref.\,\cite{noi}), compared to 
lattice data taken from Ref.~\cite{Allton:2003vx}.}
\label{fig5}}
\end{figure}

At non-zero chemical potential, quantities of interest that have become 
accessible in lattice QCD are the ``pressure difference" and the quark number 
density.
The (scaled) pressure difference is defined as:
\bea
\frac{\Delta p\left(T,\mu\right)}{T^4}=\frac{p\left(T,\mu\right)-p\left(T,
\mu=0\right)}{T^4}.
\eea
A comparison of $\Delta p$, calculated in the PNJL model, with two-flavour lattice results 
is presented in Fig.\,\ref{fig5}. This figure shows
the scaled pressure difference as a function of the temperature for a series of
chemical potentials, with values ranging between $\mu=0.2\,T_{c}^{(0)}$ and
$\mu \simeq T_{c}^{(0)}$ where $T_{c}^{(0)} \equiv T_c(\mu = 0)$. The agreement between our results \cite{noi} and the lattice 
data is quite satisfactory.
\begin{figure}
\hspace{-.05\textwidth}
\begin{minipage}[t]{.48\textwidth}
\includegraphics*[width=\textwidth]{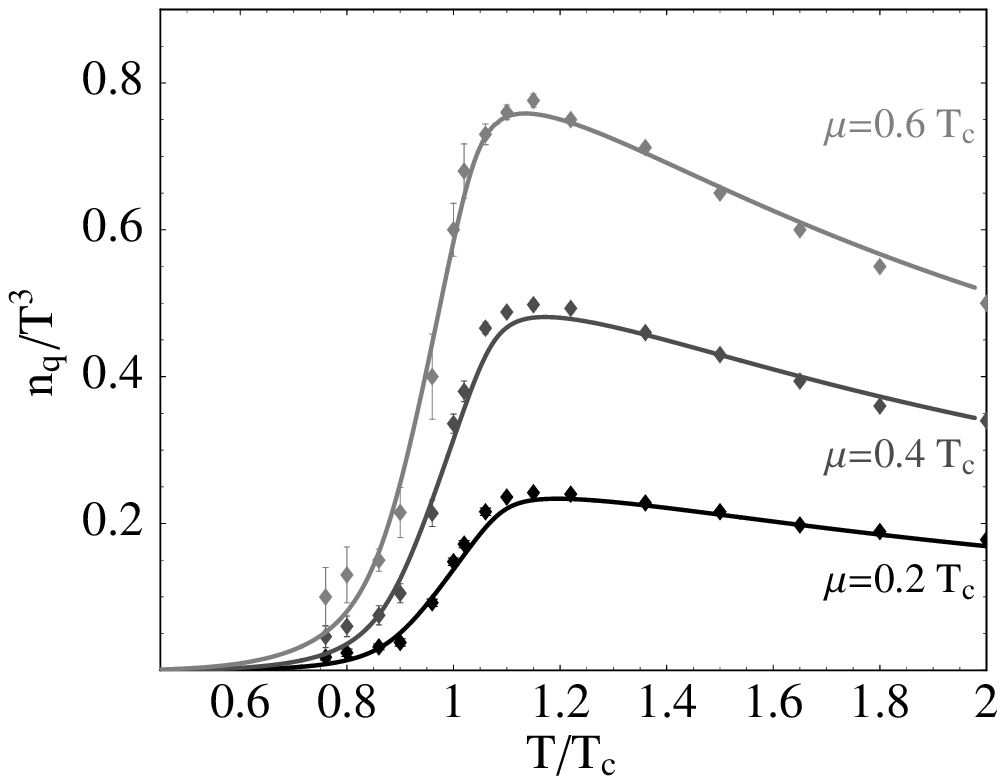}\\
\centerline{(a)}
\end{minipage}
\hspace{.02\textwidth}
\begin{minipage}[t]{.48\textwidth}
\includegraphics*[width=\textwidth]{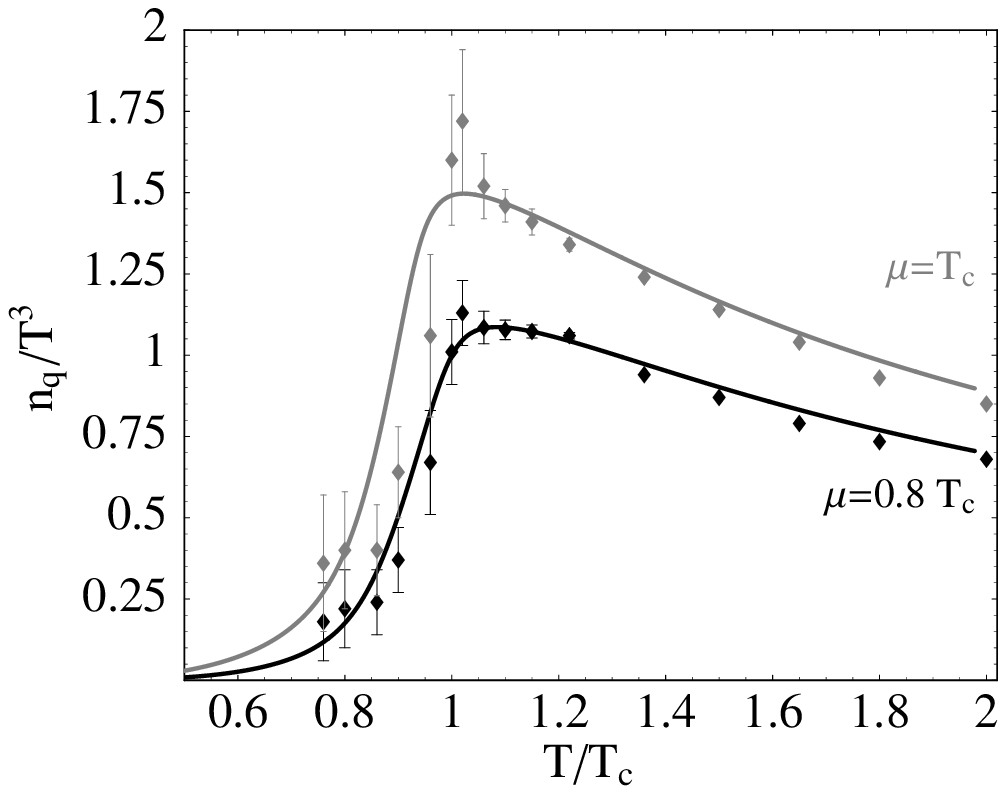}\\
\centerline{(b)}
\end{minipage}
\parbox{15cm}{
\caption{
\footnotesize Scaled quark number densities \cite{noi} as a function of temperature at
different values of the chemical potential, compared to 
lattice data taken from Ref.~\cite{Allton:2003vx}.}
\label{fig6}}
\end{figure}

A related quantity for which lattice results at finite $\mu$ exist, is
the scaled quark number density, defined as:
\beq
\frac{n_q\left(T,\mu\right)}{T^3}=-\frac{1}{T^3}\frac{\partial\Omega
\left(T,\mu\right)}{\partial\mu}.
\eeq
Results \cite{noi} for $n_q$ as a function of the temperature,
for different values of the quark chemical potential, are shown in Fig.\,\ref{fig6}
in comparison with corresponding lattice data \cite{Allton:2003vx}. Also in this case, the 
agreement between our PNJL model  and the corresponding lattice
data is surprisingly good. 

It is a remarkable feature that the quark densities and the pressure difference
at finite $\mu$ are
so well reproduced even though the lattice ``data'' have been obtained by a 
Taylor expansion up to fourth
order in $\mu$, whereas our thermodynamical potential is used with its full 
functional dependence on $\mu$. We have examined the convergence in powers of 
$\mu$ by expanding Eq.\,(\ref{omega2}). It turns out that the Taylor expansion to
 order $\mu^2$ deviates from the full result by less than 10 \% even at a 
chemical potential as large as $\mu\sim T_c$. When expanded to 
${\cal O}(\mu^4)$,
no visible difference is left between the approximate and full calculations for
all cases shown in Figs.~\ref{fig5} and~\ref{fig6}.

An exact copy of our PNJL model \cite{noi} has recently been 
employed in \cite{GMMR06} to calculate susceptibilties and higher order derivatives in the expansion of the pressure $p(T,\mu) = -\Omega(T,\mu)$ around $\mu = 0$. 
\begin{equation}
{p(T,\mu)\over T^4}=\sum_{n~even} c_n(T)\left({\mu\over T}\right)^n~~.
\end{equation}
The resulting quark number susceptibility
\begin{equation}
c_2(T) = {1\over 2T^2}\left({\partial^2p\over\partial\mu^2}\right)_{\mu=0}
\end{equation}
compares well with lattice QCD computations. The higher-order coefficients $c_{4,6}$ reproduce the
corresponding lattice data around $T_c$ very well, but $c_4$ as obtained in the PNJL calculation tends to be too large at higher temperatures. For a more quantitative understanding, further steps are yet  necessary towards a consistent treatment beyond the mean-field level \cite{RRW}.

\section{Phase diagram}
Lattice data for the QCD phase diagram exist up to relatively high 
temperatures, but extrapolations to non-zero chemical potential are still 
subject to large uncertainties. It is nonetheless instructive to explore the 
phase diagram as calculated in the PNJL model \cite{noi} in comparison with 
present lattice QCD results. In particular, questions about the sensitivity of 
this phase diagram with respect to changes of the input quark masses will be 
addressed. This is an important issue, given the fact that most lattice QCD 
computations so far encounter technical limitations which restrict the input 
bare quark masses to relatively large values. The PNJL approach permits to
vary the bare quark mass in a controlled way compatible with explicit chiral 
symmetry breaking in QCD. One can therefore interpolate between 
large quark masses presently accessible in lattice simulations, the physically 
relevant range of light quark masses around 5 MeV and further down to the 
chiral limit.
\begin{figure}
\hspace{.2\textwidth}
\parbox{5cm}{
\scalebox{.95}{
\includegraphics*[width=.55\textwidth]{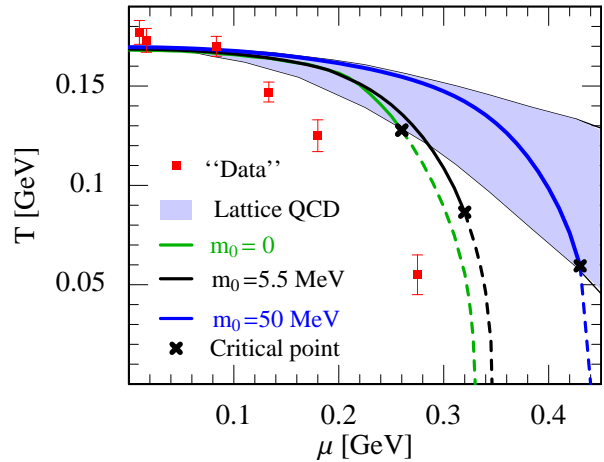}\\
}}\\
%
%
\parbox{15cm}{
\caption{
\footnotesize Phase diagram of the PNJL model for different values of the
bare quark mass $m_0$. Dashed lines correspond to a first order phase transition,
full lines to a crossover. The band represents the extrapolation from lattice QCD results
\cite{Allton:2002zi}. Also shown for orientation are the chemical freezeout ``data'' obtained through a thermal fit~\cite{andronic}.}
\label{fig7}}
\end{figure}

Fig.\,\ref{fig7} presents our two-flavour PNJL results of the phase 
boundaries in the $(T,\mu)$ plane. These calculations
should still be considered as an exploratory study since they do not yet include explicit diquark degrees of freedom, an important ingredient when turning to large chemical potentials, and the 
Polyakov loop fields are still treated in an approximate mean-field framework. Some interesting tendencies are, however, already apparent at the present stage.
 
\begin{figure}
\hspace{-.05\textwidth}
\begin{minipage}[t]{.48\textwidth}
\parbox{5cm}{
\scalebox{.95}{
\includegraphics*[width=\textwidth]{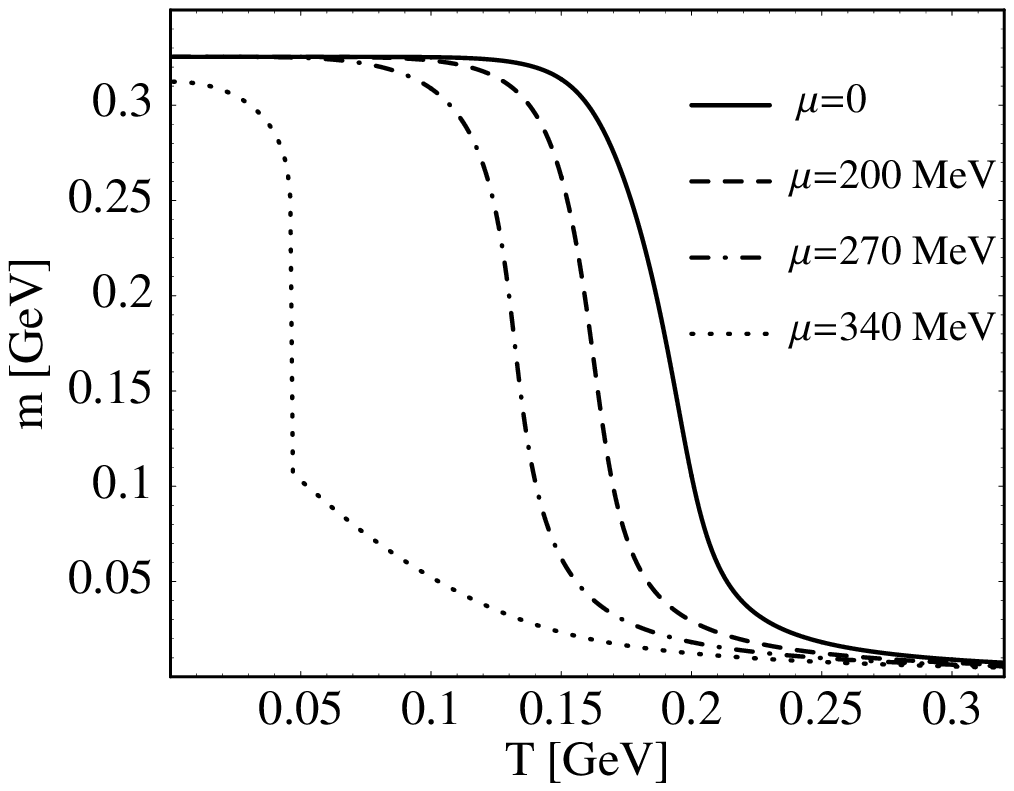}\\}}
\centerline{(a)}
\end{minipage}
\hspace{.02\textwidth}
\begin{minipage}[t]{.48\textwidth}
\parbox{5cm}{
\scalebox{0.95}{
\includegraphics*[width=\textwidth]{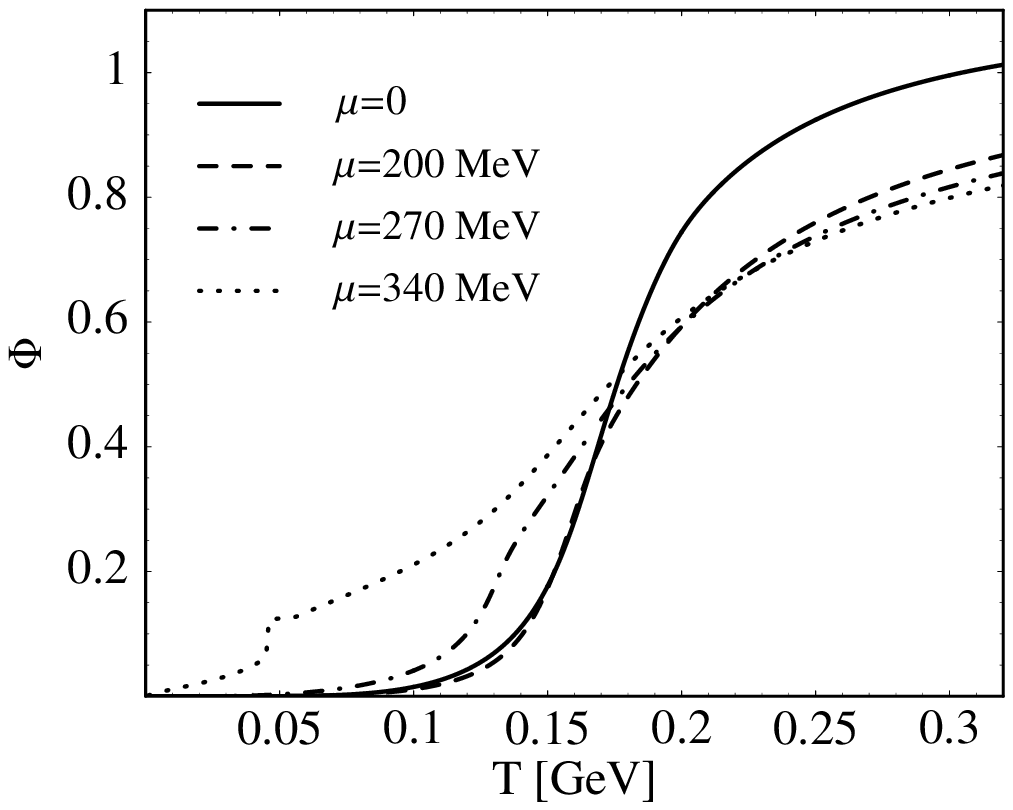}\\}}
\centerline{(b)}
\end{minipage}
\parbox{15cm}{
\caption{
\footnotesize Constituent quark mass $(a)$ and Polyakov loop
$(b)$ as functions of temperature for different values of the chemical 
potential. In both figures, $m_0=5.5$ MeV.}
\label{fig8}}
\end{figure}

Curves are shown for three different values of the bare quark mass. For $m_0=50$ MeV 
our result falls within the broad band of lattice extrapolations using an expansion 
in powers of the quark chemical potential. Reducing the bare quark
mass toward physically realistic values leaves the phase diagram at small chemical potentials basically unchanged. However, the phase boundary is shifted quite significantly to lower temperatures at increasing chemical potentials when $m_0$ is lowered.
Also shown in the figure is the position of the critical point separating 
crossover from first order phase transition. In fact, examining the
chiral condensate and the Polyakov loop as functions of temperature for
a broad range of chemical potentials (see Fig.\,\ref{fig8}), one observes that 
there is a critical chemical potential above which these two quantities indicate a discontinuous 
jump from the confined (chirally broken)  to the deconfined 
(chirally restored) phase. While a more precise location of this critical point is subject to further refined calculations \cite{RRW}, the qualitative features outlined here are expected to remain, such as the observation that  the position of the critical point depends sensitively on the input quark mass. 
\section{Conclusions}
The PNJL approach represents a minimal synthesis of the two basic principles 
that govern QCD at low temperatures: spontaneous chiral symmetry breaking and 
confinement. The respective order parameters (the chiral quark condensate and 
the Polyakov loop) are given the meaning of collective degrees of freedom. 
Quarks couple to these collective fields according to the symmetry rules 
dictated by QCD itself.

A limited set of input parameters is adjusted to reproduce lattice QCD results in the pure 
gauge sector and pion properties in the hadron sector. Then the quark-gluon 
thermodynamics above $T_c$ up to about twice the critical temperature is well 
reproduced, including quark densities up to chemical potentials of about 0.2 
GeV. In particular, the PNJL model correctly describes the step from the 
first-order deconfinement transition observed in pure-gauge lattice QCD (with 
$T_c \simeq 270$ MeV) to the crossover transition (with $T_c$ around 200 MeV) 
when $N_f = 2$ light quark flavours are added. The non-trivial result is that 
the crossovers for chiral symmetry restoration and deconfinement almost
coincide, as found in lattice simulations. The model also reproduces the quark 
number densities and pressure difference at various chemical potentials 
surprisingly well when confronted with corresponding lattice data. 
Considering that the lattice results have been found by a Taylor expansion in 
powers of the chemical potential, this agreement indicates rapid convergence of 
the power series in $\mu$.

The phase diagram predicted in this model has interesting implications. 
Starting from large quark masses an extrapolation 
to realistic small quark masses can be performed. The location of the critical point turns 
out to be sensitive to the input value of the bare (current) quark mass. 

The conclusion to be drawn at this point is as follows. A quasiparticle approach, with its dynamics rooted in spontaneous chiral symmetry breaking and confinement and with parameters controlled by a few known 
properties of the gluonic and hadronic sectors of the QCD phase diagram, can 
account for essential observations from two-flavour $N_c = 3$ lattice QCD 
thermodynamics up to about twice the critical temperature of about 0.2 GeV. 
Presently ongoing further developments include:
\begin{itemize}
\item{systematic steps beyond the mean-field approximation;}
\item{extensions to 2+1 flavours;}
\item{inclusion of explicit diquark degrees of freedom and investigations of colour superconductivity in the high density domain;} 
\item{detailed evaluations of susceptibilities and transport properties at finite chemical potential.}
\end{itemize} 


\begin{thebibliography}{10}

\bibitem{NJL61}
Y. Nambu and G. Jona-Lasinio,
Phys. Rev. {\bf 122}, 345 (1961).

\bibitem{VW91}
U. Vogl and W. Weise,
Prog. Part. Nucl. Phys. {\bf 27}, 195 (1991).

\bibitem{Kl92}
S.P. Klevanski,
Rev. Mod. Phys. {\bf 64}, 3 (1992).

\bibitem{HK94}
T. Hatsuda and T. Kunihiro,
Phys. Rep. {\bf 247}, 221 (1994).

\bibitem{Ri97}
G. Ripka,
{\it Quarks Bound by Chiral Fields}, Clarendon, Oxford (1997).

\bibitem{Bu05}
M. Buballa,
Phys. Rep. {\bf 407}, 205 (2005).

\bibitem{Barducci:2005ut}
  A.~Barducci, R.~Casalbuoni, G.~Pettini and L.~Ravagli,
  Phys.\ Rev.\ D {\bf 72}, 056002 (2005).

\bibitem{ARW98}
M. Alford, K. Rajagopal and F. Wilczek, Phys. Lett. {\bf B422}, 247 (1998);
R. Rapp, T. Sch\"afer, E.V. Shuryak and M. Velkovsky, Phys. Rev. Lett. {\bf 81}, 53 (1998).

\bibitem{DiG}
A. Di Giacomo, H.G. Dosch, V.I. Shevchenko and Y.A. Simonov,
Phys. Rep. {\bf 372}, 319 (2002).

\bibitem{Ratti}
C. Ratti and W. Weise,
Phys. Rev. {\bf D70}, 054013 (2004).

\bibitem{noi}
C. Ratti, M. A. Thaler and W. Weise,
Phys. Rev. {\bf D73}, 014019 (2006).

\bibitem{Fukushima:2003fw}
K.~Fukushima,
\newblock Phys. Lett. B {\bf 591}, 277 (2004).

\bibitem{DPZ05}
A. Dumitru, R.D. Pisarski and D. Zschiesche,
Phys. Rev. {\bf D72}, 065008 (2005).

\bibitem{RRW}
S. R\"o{\ss}ner, Diploma Thesis (2006);
S. R\"o{\ss}ner, C. Ratti and W. Weise, to be published.

\bibitem{Meisinger:2003id}
  P.~N.~Meisinger, M.~C.~Ogilvie and T.~R.~Miller,
  Phys.\ Lett.\ B {\bf 585}, 149 (2004).

\bibitem{Boyd}
G.~Boyd {\em et~al.},
\newblock Nucl. Phys. B {\bf 469}, 419 (1996).

\bibitem{Kaczmarek:2002mc}
O.~Kaczmarek, F.~Karsch, P.~Petreczky, and F.~Zantow,
\newblock Phys. Lett. B {\bf 543}, 41 (2002).

\bibitem{Karsch}
F.~Karsch,
\newblock Lecture Notes in Phys. (Springer) {\bf 583}, 209 (2002);\\
F.~Karsch, E.~Laermann and A.~Peikert,
\newblock Nucl. Phys. B {\bf 605}, 579 (2002).

\bibitem{AliKhan:2001ek}
A.~Ali Khan {\it et al.}  
Phys.\ Rev. D {\bf 64}, 074510 (2001).

\bibitem{Allton:2003vx}
C.~R. Allton {\em et~al.},
\newblock Phys. Rev. D {\bf 68}, 014507 (2003).

\bibitem{GMMR06}
S.K. Ghosh, T.K. Mukherjee, M.G. Mustafa and R. Ray,
hep-ph/0603050 (2006).

\bibitem{Allton:2002zi}
  C.~R.~Allton {\it et al.},
  Phys.\ Rev.\ D {\bf 66}, 074507 (2002).

\bibitem{andronic}
A. Andronic and P. Braun-Munzinger,
Lect. Notes Phys. {\bf 652}, 35 (2004).

\end{thebibliography}
\end{document}